\documentclass[10pt,letterpaper]{article}
\usepackage{opex3}
\usepackage{amsmath}
\usepackage{cite}

\renewcommand{\vector}[1]{\mathbf{#1}}
\newcommand{\eh}[0]{\eta_H}
\newcommand{\ea}[0]{\eta_A}
\newcommand{\eb}[0]{\eta_B}
\newcommand{\dch}[0]{d_H}
\newcommand{\dca}[0]{d_A}
\newcommand{\dcb}[0]{d_B}
\newcommand{\gt}[0]{g^{(2)}(0)}
\newcommand{\mmat}[0]{M_{\eta}}

\begin{document}

\title{Fast and simple characterization of a photon pair source}

\author{F.~Bussi\`eres$^{1,2,\dag}$, J.~A.~Slater$^{2,\dag}$, N.~Godbout$^{1}$, and W.~Tittel$^{2}$}

\address{$^{1}$ Laboratoire des fibres optiques, D\'epartement de g\'enie physique, \'Ecole Polytechnique de Montr\'eal, CP~6079, succ. Centre-Ville, Montr\'eal (Qu\'ebec), H3C~3A7 Canada. \\
$^{2}$ Institute for Quantum Information Science, Department of Physics and Astronomy, University of Calgary, Calgary (Alberta), T2N~1N4 Canada.}

\email{jslater@qis.ucalgary.ca} 

\begin{abstract}
We present an exact model of the detection statistics of a probabilistic source of photon pairs from which a fast, simple and precise method to measure the source's brightness and photon channel transmissions is demonstrated. We measure such properties for a source based on spontaneous parametric downconversion in a periodically poled LiNbO$_3$ crystal producing pairs at 810 and 1550~nm wavelengths. We further validate the model by comparing the predicted and measured values for the $g^{(2)}(0)$ of a heralded single photon source over a wide range of the brightness. Our model is of particular use for monitoring and tuning the brightness on demand as required for various quantum communication applications. We comment on its applicability to sources involving spectral and/or spatial filtering.
\end{abstract}



\section{Introduction}

Sources of photon pairs are an essential building block in implementations of Quantum Communication protocols. Examples of such are Quantum Key Distribution (QKD), enabling unconditional security in the exchange of confidential messages~\cite{BB84,Ekert91}, or Quantum Repeaters, needed to break the distance barrier of QKD~\cite{BDCZ98,DLCZ01}. Photon pairs obtained from Spontaneous Parametric Downconversion (SPDC)~\cite{BW70} or Spontaneous Four-Wave-Mixing (SFWM)~\cite{FVSK02} in nonlinear materials, or from atomic ensembles~\cite{K03,STTV07}, can be used to generate either entangled photons by a careful arrangement of two downconversion paths~\cite{TW01}, or to create a single photon source by announcing the presence of one photon through the detection of the other one (a Heralded Single Photon Source, or HSPS)~\cite{HM86}.

All aforementioned sources are of probabilistic nature, i.e. the number of emitted pairs per time unit follows a given statistical distribution such as a Poissonian or thermal distribution. In most applications, it is beneficial or even essential to know the mean number of photon pairs $\mu$ emitted per unit of time, a quantity that we shall refer to as the \textit{brightness}. For entanglement based QKD, Ma~\textit{et al.} have shown that both the key generation rate and the maximum distance over which a secret key can be established can be maximized by properly tuning the brightness~\cite{MFL07}. Another example is the security of QKD based on HSPS, which relies on the ability of the sender to assess the photon statistics in a precise way~\cite{WSY02,Chen08,Adachi07,MS07}. Also, de~Riedmatten~\textit{et al.} have shown that the visibility in Bell-state measurements, which is a key element of proposed quantum repeaters, crucially depends on the brightness~\cite{RMTZG03}.

Assessing the brightness of a source of photon pairs is a non-trivial task when limited to lossy channels and non photon-number-resolving detectors. This problem can be solved provided one knows the exact value of the total transmission of all photon channels. However, evaluating the loss associated with coupling a single photon from free-space to a singlemode fibre is not simple. One technique requires mode-matching a probe laser to the single photon mode, but this can be imprecise and unpractical (see~\cite{TOS04,OTS05,TL07} as examples). The brightness can also be inferred from measurements of the second-order autocorrelation function, $\gt$~\cite{Mandel95}. However, as the time required for $\gt$ measurements depends on three-fold coincidence detection stemming from two simultaneously generated pairs, such measurements are time consuming to implement (see~\cite{STTV07} and \cite{MRTSZG02} as examples). Therefore, a method from which the brightness and the losses of the transmission lines can be determined with precision, speed and simplicity is necessary.

In this work, we show how one can assess the brightness and the photon channel transmissions of a source of photon pairs by solely measuring single and two-fold coincidence detections stemming from photons belonging to one pair. This makes this method very fast and efficient. In section~\ref{section:model}, we present a model describing the detection statistics of any probabilistic source of photon pairs, and we show how the brightness and the losses of the channels can be assessed. Then, in section~\ref{section:results}, we describe an implementation of the proposed method and use it to predict the value of the autocorrelation function, $\gt$, of a HSPS for a wide range of brightness values. We confirm the model by the direct measurement of the predicted $\gt$ values. Finally, in section~\ref{section:correlations}, we discuss the limits of our model when the generated photons are spectrally correlated.

\section{An exact model of the detection statistics} \label{section:model}
\subsection{Description of the model} \label{subsection:modeldescrip}
\begin{figure}[htbp]
\centering\includegraphics[width=7cm]{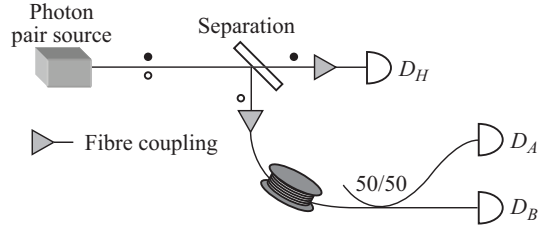}
\caption{The sources of photon pairs we consider comprise all probabilistic sources, including those based on nonlinear crystals, optical fibres or atomic ensembles. The distribution of the number of produced photon pairs per measurement time window can be given by any distribution such as Poissonian or thermal and is assumed to be known in advance. The pairs are deterministically separated, potentially by a dichroic beamsplitter in the case of collinear generation with non-degenerate wavelengths, or by non-collinear generation, into two separate channels. Each beam is filtered to remove all pump light and then the pairs are coupled into optical fibres. One beam is split again at a 50/50 beamsplitter before the photons are detected by non-photon number resolving single photon detectors $D_H$, $D_A$ and $D_B$.} \label{fig:genericsetup}
\end{figure}

To assess the properties of a source of photon pairs, we developed an exact model of the detection statistics of the experimental setup detailed in Fig.~\ref{fig:genericsetup}.

To model the detection statistics of this experimental setup we construct a column vector $\vector{P}$, as shown in Eq.~(\ref{state-vector-def}), which describes the joint state of the detectors:
\begin{equation}
\label{state-vector-def}
\vector{P} =
\left( \begin{smallmatrix}
  p_{\bar{A}\bar{B}\bar{H}} & p_{A\bar{B}\bar{H}} & p_{\bar{A}B\bar{H}} & p_{\bar{A}\bar{B}H} & p_{AB\bar{H}}
  &  p_{A\bar{B}H} & p_{\bar{A}BH} & p_{ABH} \\
\end{smallmatrix} \right)^{\mathrm{T}}.
\end{equation}
Each element of $\vector{P}$ describes the probability that a set of detectors clicked or not per measurement time window, which is defined as the elementary observation time (e.g. a short time window centered on one pump pulse; see later) for which detections are considered for statistical analysis. For example, $p_{A\bar{B}\bar{H}}$ is the probability that detector $D_A$ clicked, during the measurement time window, and $D_H$ and $D_B$ did not. The goal is to determine how this vector, initially in state $\vector{P}_0 = \left( \begin{smallmatrix} 1 & 0 & \ldots & 0 \end{smallmatrix} \right)^{\text{T}}$, is affected by single and multiple photon pair emissions as well as detector dark counts during one measurement time window. First, we describe the interaction of \textit{one} photon pair with the detectors using the following transition matrix:
\begin{equation}
\mmat =
\left(
  \begin{smallmatrix}
    1 - \eh + (\ea + \eb)(\eh -1) & 0 & 0 & 0 & 0 & 0 & 0 & 0 \\
    \ea(1-\eh) & (1 - \eb)(1 - \eh) & 0 & 0 & 0 & 0 & 0 & 0 \\
    \eb(1-\eh) & 0 & (1 - \ea)(1 - \eh) & 0 & 0 & 0 & 0 & 0 \\
    \eh( 1 - (\ea + \eb)) & 0 & 0 & 1 - (\ea + \eb) & 0 & 0 & 0 & 0 \\
    0 &\eb(1-\eh) & \ea(1-\eh) & 0 & 1-\eh & 0 & 0 & 0 \\
    \ea \eh & \eh(1 - \eb) & 0 & \ea & 0 & 1-\eb & 0 & 0 \\
    \eb \eh & 0 & \eh(1 - \ea) & \eb & 0 & 0 & 1-\ea & 0 \\
    0 & \eb \eh & \ea \eh & 0 & \eh & \eb & \ea & 1 \\
  \end{smallmatrix}
\right). \label{M-matrix-def}
\end{equation}
Each element of $\mmat$ describes the probability for a pair to cause a transition of the three detectors. Each term is written as a function of $\eh$, $\ea$ and $\eb$ which are the overall transmissions of each channel, from the photon pair source to $D_H$, $D_A$ and $D_B$ respectively, including all optical losses, fibre coupling losses, detector inefficiencies, and the 50/50 beamsplitter. For example, $\mmat(1,1)$ is the probability for the system to make a transition from $\bar{A}\bar{B}\bar{H}$ to $\bar{A}\bar{B}\bar{H}$ (i.e. to remain in the state where no detectors clicked), which must equal $p_{\bar{A}\bar{H}} + p_{\bar{B}\bar{H}} = (1 - \eh - \ea + \ea\eh) + (1 - \eh - \eb + \eb\eh) = 1 - \eh + (\ea + \eb)(\eh -1)$. Similarly, $\mmat(2,1)$ is the probability to make a transition from $\bar{A}\bar{B}\bar{H}$ to $A\bar{B}\bar{H}$ (i.e. no detectors clicked before and, provided one photon pair arrives, only $D_A$ clicks), which equals $\ea(1-\eh)$. All the upper diagonal elements are equal to~0 as photons cannot make detectors ``unclick''. The rest of the matrix is constructed following the same physical reasoning. Furthermore, to conserve the total probability, each column of $\mmat$ sums to~$1$. The result of \textit{one} photon pair interacting with the detectors is thus given by $\mmat\vector{P}_0$.

Second, the evolution of the system when~$i$ photon pairs are created during the measurement time window is described by $(\mmat)^i\vector{P}_0$, as losses and the beamsplitter choice for individual pairs in multi-pair emission are independent.

In addition to the absorption of a photon, thermal excitations can also cause detector clicks. These dark counts can be taken into account by constructing another matrix $M_{dc}$. Thus, the evolution resulting from dark counts and $i$ photon pairs is described by $M_{dc}(\mmat)^i\vector{P}_0$. Noting the dark count probabilities per measurement time window as $d_H$, $d_A$ and $d_B$, we get
\begin{equation}
M_{dc} =
\left(
  \begin{smallmatrix}
    (1-\dca)(1-\dcb)(1-\dch) & 0 & 0 & 0 & 0 & 0 & 0 & 0 \\
    \dca(1-\dcb)(1-\dch) & (1-\dcb)(1-\dch) & 0 & 0 & 0 & 0 & 0 & 0 \\
    (1-\dca)\dcb(1-\dch) & 0 & (1-\dca)(1-\dch) & 0 & 0 & 0 & 0 & 0 \\
    (1-\dca)(1-\dcb)\dch & 0 & 0 & (1-\dca)(1-\dcb) & 0 & 0 & 0 & 0 \\
    \dca \dcb (1-\dch) & \dcb(1-\dch) & \dca(1-\dch) & 0 & 1-\dch & 0 & 0 & 0 \\
    \dca(1-\dcb)\dch & (1-\dcb)\dch & 0 & \dca(1-\dcb) & 0 & 1-\dcb & 0 & 0 \\
    (1-\dca)\dcb\dch & 0 & (1-\dca)\dch & (1-\dca)\dcb & 0 & 0 & 1-\dca & 0 \\
    \dca \dcb \dch & \dcb \dch & \dca \dch & \dca\dcb & \dch & \dcb & \dca & 1 \\
  \end{smallmatrix}
\right). \label{matrix-dc}
\end{equation}

Thus, when an unknown number of photon pairs are incident, it is possible to calculate the final vector $\vector{P}$ through
\begin{equation}
\vector{P} =
\sum_{i=0}^{\infty} p_i M_{dc}(M_{\eta})^i \vector{P}_0,
\label{eqn:master-matrix}
\end{equation}
where $p_i$ is the probability to create $i$ photon pairs per measurement time window. Provided that the probability distribution for $p_i$ is known this equation holds for all distributions, such as Poissonian, thermal or any distributions between the two~\cite{RSMATZG04}. Note that all matrices commute so the order in which they are applied does not matter. The construction of the matrices ensures that all elements of $\vector{P}$ are bounded individually between~0 and~1 and that the elements of $\vector{P}$ sum to~1, i.e. the total probability is conserved. We note that the model is exact and there are no approximations.

\subsection{Determining the photon channel transmissions}
\begin{figure}[!h]
\centering\includegraphics[width=7cm]{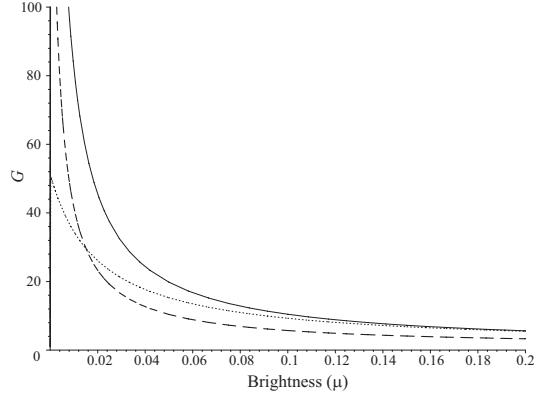}
\caption{Correlation strength $G$ versus brightness $\mu$. The solid line corresponds to $\eh = 60\%$ and $\ea = 25\%$. It reaches a maximum value at very low $\mu$ and then sharply decreases to~1 for $\mu = 0$ (not visible). The meanings of the dotted and dashed lines are discussed in section~\ref{section:results} and~\ref{section:correlations} respectively.} \label{fig:G}
\end{figure}

We now show how one can determine precisely the values of $\mu$, $\eh$, $\ea$ and $\eb$ by measuring single and two-fold coincidence detection probabilities stemming from single pairs only. For these measurements, we require that the pump power (or equivalently the brightness of the photon pair source) is low enough so that multi-pair events are negligible: $p_i \ll p_1$ for $i>1$. Experimentally, this can be verified by looking at how correlated detections on $D_A$ are to detections on $D_H$. To measure this, we first define $p_H$ to be the heralding probability, i.e. the probability for $D_H$ to click independent of the other detectors, $p_H = p_{\bar{A}\bar{B}H} + p_{A\bar{B}H} + p_{\bar{A}BH} + p_{ABH}$, and the similar expressions for $p_{AH}$ and $p_{A}$. We then define a parameter $G = p_{AH}/(p_{H}p_{A})$ quantifying the strength of the correlation between detections at $D_A$ and $D_H$. The model described by Eq.~(\ref{eqn:master-matrix}) predicts that, for Poisson, thermal and in between distributions, the value of $G$ equals one at a very low brightness, when the coincidences are dominated by dark counts and detections are uncorrelated, and equals one again at high heralding probabilities, when the coincidence detections stem mostly from multi-pair emissions and correlations are smeared out. In between, the value of $G$ can go well above~1 and this is an indication that multi-pair emissions are negligible.  As we show here, this allows to experimentally obtain an upper bound for $\mu$ when proceeding as follows. First, the dark count probability per measurement time window for each detector is measured. Second, the pump power is lowered and the transmissions are optimized until a value of $G$ significantly higher than~1 is measured. Third, a plot of $G$ versus $\mu$ is produced numerically assuming that the fibre coupling is perfect and that there are no additional optical losses, thereby setting the values of $\eh$ and $\ea$ equal to the specified detection efficiency of the detectors. Finally, an upper bound for $\mu$ is obtained from this plot by identifying the largest value of $\mu$ that produces a value of $G$ equal to the measured value. They key point is that, for a given $\mu$ and dark count probabilities, $G$ is decreased towards~1 when the transmissions are decreased. Thus, using this method, the true value of $\mu$ must be smaller than the upper bound as the transmissions are overestimated. This, in return, allows one to obtain a lower bound for the ratio, $r = p_1/p_{i>1}$,  of the probability of single pair emissions, $p_1$, over the probability of multi-pair emissions, $p_{i>1} = 1 - p_0 - p_1$. As an illustration, using $\eh = 60\%$ and $\ea = 25\%$, corresponding to the detection efficiencies of our detectors, and using their respective measured dark count probabilities (see section~\ref{section:results}), we produced the solid line shown on Fig.~\ref{fig:G} where we assumed a Poisson distribution, $p_i = \exp(-\mu)\mu^i/i!$.

Once the pump power is properly set and the lower bound on $r$ is satisfactory, Eq.~(\ref{eqn:master-matrix}) can be truncated to $i=1$ and one can show that the probability for $D_H$ to click on a photon and not a dark count is given by $p_{H}^{(1)} = (p_{H} - d_H)/(1 - d_H)$. Similarly, we get \mbox{$p_A^{(1)} = (p_A - d_A)/(1-d_A)$} and the equivalent for $p_B^{(1)}$. In the same way, we can get expressions for the coincidence probabilities $p_{AH}$ and $p_{BH}$. Then, using these expressions and an experimental data collection run with a heralding probability that guarantees negligible multi-pair events, one can solve for the four unknowns $\mu$, $\eh$, $\ea$ and $\eb$, since the dark count probabilities can be measured directly. These unknowns can be calculated through equations~(\ref{eqn:eh-calc}) through~(\ref{eqn:mu-calc}). The equivalent set for $D_B$ is constructed by replacing $\ea$ and $p_A^{(1)}$ by $\eb$ and $p_B^{(1)}$, respectively:
\begin{equation}
\eh = \dfrac{p_{AH} - p_H^{(1)} d_A (1-d_H) - p_A^{(1)}d_H(1-d_A) - d_A d_H}{p_A^{(1)}(1-d_A)(1-d_H)},
\label{eqn:eh-calc}
\end{equation}
\begin{equation}
\label{eqn:ea-calc}
\ea = \dfrac{p_{AH} - p_H^{(1)}d_A(1- d_H) - p_A^{(1)}d_H(1-d_A) - d_Ad_H}{p_H^{(1)}(1-d_A)(1-d_H)},
\end{equation}
\begin{equation}
\label{eqn:mu-calc}
p_1 = \dfrac{p_{H}^{(1)}}{\eh} = \dfrac{p_{A}^{(1)}}{\ea}.
\end{equation}
Note that these predictions apply to any statistical distribution for which the multi-pair events can be neglected (for example, through the method described above). However, to determine the value of the brightness, one must have prior knowledge of the distribution and how to relate it to the measured value of~$p_1$. In the case of a Poissonian source, we have $p_1 = \mu \exp(-\mu)$ which can be solved numerically for $\mu$. The case of a thermal distribution is similar with \mbox{$p_1 = (\tanh \mu/\cosh \mu)^2$}.

Once the transmissions are precisely known, one can use Eq.~(\ref{eqn:master-matrix}) to find the brightness that corresponds to any measured heralding probability. This will then allow to predict the complete detection statistics vector $\vector{P}$.

\subsection{Application to a HSPS}
The transmissions and the brightness, along with the knowledge of the pair distribution type, can be used to predict the second-order autocorrelation function of the heralded mode of a HSPS, $\gt$, for any desired heralding probability $p_H$, in a Hanbury Brown and Twiss (HBT) experiment~\cite{HBT56}. The distribution of the number of photons in that mode follows the distribution of the number of photon pairs created by the source except for a reduced vacuum component, $p_0$, due to the heralding. A $\gt < 1$, which is achievable with a HSPS, implies a nonclassical source (for a perfect single photon source $\gt = 0$).  Alternatively, a $\gt \geq 1$ describes a classical source (for Poissonian \mbox{$\gt = 1$} and for thermal $\gt = 2$).
 To verify our model, we compare its predictions with a real measurement of the $\gt$. In this experiment, which can be seen as measuring a subset of Eq.~(\ref{eqn:master-matrix}), detectors $D_A$ and $D_B$ are activated only when $D_H$ clicks. The $\gt$ is defined as
\begin{equation}
  \gt = \frac{p_{AB | H}}{p_{A | H} \times p_{B | H}},
\end{equation}
where $p_{AB|H}$ is the probability that both $D_A$ and $D_B$ click provided that $D_H$ clicked, etc.

For a specific heralding probability $p_H$, we can directly measure $\gt$ using the setup of Fig.~\ref{fig:genericsetup} by keeping only the events where $D_H$ clicked. On the other hand, the $\gt$ can also be predicted for the same heralding probability using Eq.~(\ref{eqn:master-matrix}). The experimental results of this verification are presented in the next section.

One interesting result can be derived from our model. Considering a Poissonian distribution at low brightness and assuming that dark counts are negligible, one can derive from Eq.~(\ref{eqn:master-matrix}) that $\gt = \mu(2-\eh)$~\cite{thermalg2}. This indicates that for a HSPS, the transmission to the heralding detector is a crucial parameter to optimize.

It is important to note here that the 50/50 beamsplitter used in the setup of Fig.~\ref{fig:genericsetup} is not required to determine the brightness and the transmissions. Indeed, to assess the brightness and the transmissions only the detectors $D_H$ and $D_A$ are necessary. In this work, the beamsplitter and $D_B$ were added only to provide a way to verify the validity of the predictions through the $\gt$ measurement.  Modifying the vector $\vector{P}$ and the matrices to accommodate for a setup with no beamsplitter and $D_B$ is straightforward.

\section{Experimental results} \label{section:results}
\begin{figure}[!htb]
\centering\includegraphics[width=7cm]{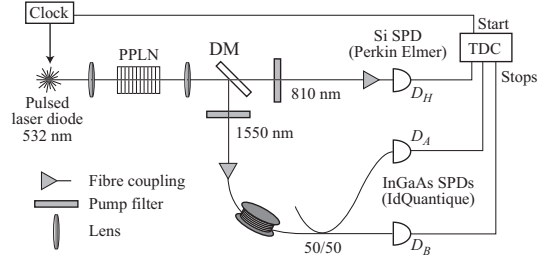}
\caption{Experimental setup.} \label{fig:realsetup}
\end{figure}

The experimental setup is shown in Fig.~\ref{fig:realsetup}.
A clocking signal triggers a pulsed laser diode (PicoQuant PicoTA) creating 50~ps pulses at 532~nm filtered to remove excess 1064~nm light. The pulses are focused onto a 1~cm long periodically poled LiNbO$_3$ crystal (PPLN) from Stratophase with a grating period of 7.05~\textmu m heated to 175.7~$^{\circ}$C. Then, collinear spontaneous parametric downconversion to one or several photon pairs can occur, each pair consisting of one 810~nm and one 1550~nm photon. A dichroic mirror DM separates the two wavelengths and, after removing the 532~nm light with long-pass color filters, the photons are coupled into SMF28 optical fibres. The 810~nm photons are sent towards a free-running Si single photon counting module $D_H$ (SPCM-AQR-14-FC, Perkin-Elmer) and the 1550~nm photons are detected by two gated InGaAs single photon detectors $D_A$ and $D_B$ (id201, IdQuantique) positioned right after a 50/50 fibre beamsplitter. The detection statistics are recorded using a Time-Digital-Converter (TDC-GPX, ACAM), providing the time elapsed between a start pulse, given by the clock, and each stop pulse, corresponding to detections. The data is analyzed in real-time and the statistics are updated continuously until the end of each run. The coherence length $l_c$ of the downconverted 810~nm photons was measured to be $180~\mu$m using a Michelson interferometer. From these measurements we calculated the bandwidth to be $\Delta \lambda_{\text{810}} \approx 4\,\text{nm}$ and, based on energy conservation of the SPDC process, $\Delta \lambda_{\text{1550}} \approx 15\,\text{nm}$. As the downconverted photons' coherence time, which equals $l_c/c = 0.54$~ps, is much smaller than the pump pulse duration, which is $50$~ps, we can confidently assume that our source of photon pairs follows Poissonian statistics~\cite{RSMATZG04}.

The InGaAs detectors, which require gating to limit excess dark counts, were activated from each clocking signal for a 5~ns measurement time window. Detections on the Si detector were considered valid only if they arrived within a 5~ns window centered on the clocking signal, as measured by the TDC. To determine the transmissions, the clocking signal triggering the laser and InGaAs detectors was set to 30~kHz. This low repetition rate ensured that saturation effects in the detection electronics and biases in the detection statistics from the InGaAs detectors' 10~\textmu s dead-time were avoided. We first measured the dark count probabilities to be $d_A = 2.87\times 10^{-4}$, $d_B = 3.84 \times 10^{-4}$ and $d_H = 2.5 \times 10^{-7}$ per 5~ns. Next, we lowered the pump power using neutral density filters in order to increase the correlation strength between $D_H$ and $D_A$ to a value of $G = 20.6\pm 1.0$, corresponding to a heralding probability of $0.287 \pm 0.001\%$. Intersecting this value with the solid line of Fig.~\ref{fig:G} gives a upper bound of $\mu \le 0.0480 \pm 0.0013$, yielding $r \ge 41.0 \pm 2.2$, which we considered sufficiently high to continue. Next we measured the single and coincidence detection probabilities from which we obtained the following values: $\eh = 0.1212 \pm 0.0031$, $\ea = 0.0145 \pm 0.0005$, $\eb = 0.0162 \pm 0.0005$ and $\mu = 0.02375 \pm 0.00016$, corresponding to $r = 83.5 \pm 0.6$. The $G$ curve corresponding to these values is plotted as the dotted line on Fig.~\ref{fig:G}, and the predicted value of $G$ at $\mu = 0.02375$ is $23.9 \pm 0.5$, which is close to the measured value of~20.6.

\begin{figure}[htb]
\centering\includegraphics[width=13cm]{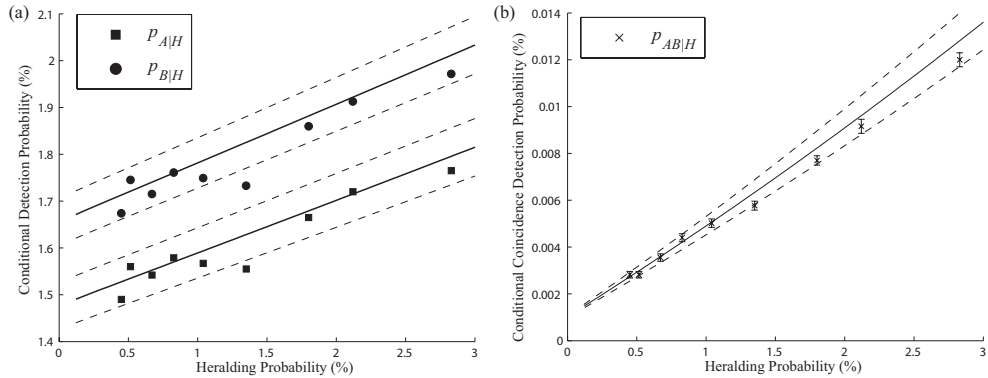}
\caption{(a) Predicted (solid lines) and measured (points) conditional detection probabilities $p_{A|H}$ and $p_{B|H}$.  (b) Predicted (solid line) and measured (points) conditional coincidence probability $p_{AB|H}$ . The dashed lines on both plots are the one standard deviation uncertainty bounds on the predicted values which were generated using the uncertainty bounds on the measured transmissions.} \label{fig:measuredprobabilities}
\end{figure}

\begin{figure}[!h]
\centering\includegraphics[width=13cm]{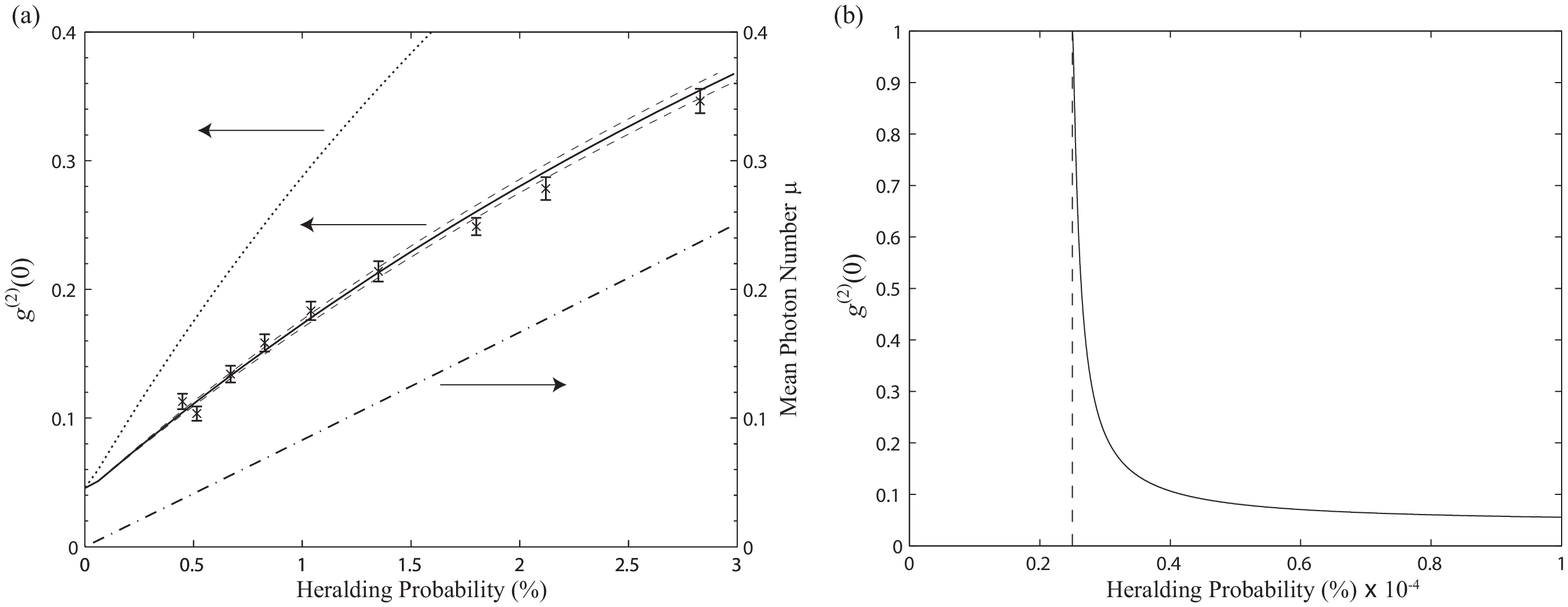}
\caption{(a) Predicted autocorrelation $\gt$ for Poissonian (solid line) and thermal (dotted line) distributions, measured values (points), and the corresponding brightness $\mu$ (dash-dotted line). The measured data agrees very well with the Poissonian distribution. The dashed lines are the
one standard deviation uncertainty bounds on the predicted values. (b) As the heralding probability reaches the noise level of $D_H$ (dashed line), the model correctly predicts that the $\gt$ approaches one, as uncorrelated dark counts begin to dominate over photon clicks.} \label{fig:g2all}
\end{figure}

Using these values together with Eq.~(\ref{eqn:master-matrix}), we produced a plot of the predicted $p_{AB|H}$, $p_{A|H}$ and $p_{B|H}$ for a wide range of the brightness (and consequently, of the heralding probability). These predictions are compared to the measured values on Fig.~\ref{fig:measuredprobabilities}(a) and~\ref{fig:measuredprobabilities}(b). Next we compared the predicted and measured $\gt$ as shown on Fig.~\ref{fig:g2all}(a). On the same figure we plotted the value of the brightness $\mu$ corresponding to each heralding probability. In all cases, the agreement between the predicted and measured values is excellent. We note that for these measurements, the repetition rate was increased to 5~MHz and the InGaAs detectors were activated for 5~ns only when the Si detector clicked synchronously (within a 5~ns window) with the pump, as required for $\gt$ measurements in the HBT setup. This resulted in an average detection rate of 30~kHz, with randomly distributed time differences, for the InGaAs detectors and was thus sufficient to ensure that saturation effects in the detection electronics was not an issue. However, to eliminate the effect of the dead-time on the detection statistics, we considered only the events where both InGaAs detectors were ready to detect photons, as provided by the ``gate out'' electrical signals of these detectors.

Our method drastically reduces the time needed to characterize the source as measurements of single and two-fold coincidence detections at a low heralding probability are sufficient to determine the transmissions. These can then be used to predict the brightness and the $\gt$ of a HSPS for any heralding probability. In contrast, a single direct measurement of the $\gt$ at a given heralding probability requires three-fold coincidence detections stemming from multi-pair emissions, which are less likely to happen. In our experiment, at a heralding probability of 0.287\%, two-fold coincidences were approximately~700 times more likely than three-fold coincidences. Consequently, a direct $\gt$ measurement required much more time.

\section{Effect of spectral and spatial correlations} \label{section:correlations}
The model proposed in section~\ref{section:model} may not apply directly to a source when spectral and/or spatial correlations exist between the photons of each pair and when the transmission of each channel are frequency and/or spatially selective. For instance, Bell state measurements generally requires filtering of photons that are spectrally correlated~\cite{BM95}. Let's suppose the spectral filtering applied on each spatially separated photons is performed with two separate filters that both need to be aligned on the photons spectra. Due to energy correlation, transmission of one photon through a spectral filter determines the spectrum of the other photon~\cite{F92}. If the filtering of the second photon does not match its now modified spectrum, the coincidence detection probability is reduced. In this case, given that one photon pair was created, we can still write the probability to get a detection at $D_H$ to be $\eh$ and at $D_A$ to be $\ea$. However, the probability to get a coincidence is lowered to $c\eh\ea$, where $0\le c \le 1$. The upper bound is reached when either the photons' spectra are uncorrelated before the filtering (see~\cite{MLS+08}), or when the selected spectra satisfy the energy conservation conditions perfectly.

When this situation applies, the detection matrix can be re-written as follows:
\begin{equation*}
M_{\eta}' =
\left(
  \begin{smallmatrix}
    1 - \eh + (\ea + \eb)(c\eh-1) & 0 & 0 & 0 & 0 & 0 & 0 & 0 \\
    \ea(1-c\eh) & 1 - \eb + \eh(c\eb - 1) & 0 & 0 & 0 & 0 & 0 & 0 \\
    \eb(1-c\eh) & 0 & 1 - \ea + \eh( c\ea - 1 ) & 0 & 0 & 0 & 0 & 0 \\
    \eh( 1 - c(\ea + \eb) ) & 0 & 0 & 1 - (\ea + \eb) & 0 & 0 & 0 & 0 \\
    0 & \eb(1-c\eh) & \ea(1-c\eh) & 0 & 1-\eh & 0 & 0 & 0 \\
    c\ea \eh & \eh(1 - c\eb) & 0 & \ea & 0 & 1-\eb & 0 & 0 \\
    c\eb \eh & 0 & \eh(1 - c\ea) & \eb & 0 & 0 & 1-\ea & 0 \\
    0 & c\eb \eh &c\ea \eh & 0 & \eh & \eb & \ea & 1 \\
  \end{smallmatrix}
\right).
\end{equation*}
The presence of spectral correlations affects the predictions of the proposed model but as we show here, the consequences are minimal. First of all, for a given brightness, a value of $c<1$ lowers the measured value of~$G$ towards~$1$. Therefore, when assessing if multi-pair events are negligible, one can assume that $c=1$ and the upper bound on $\mu$, along with the lower bound on $r$, are still valid. As an illustration, reducing~$c$ from~1, corresponding to the solid line of Fig.~\ref{fig:G}, down to~$0.5$, corresponding to the dashed line on the same figure, lowers the curve towards~$G=1$. Also, performing the above analysis to calculate the transmissions, while neglecting dark counts, we can show that the obtained solutions are $\eh' = c\eh$, $\ea' = c\ea$, $\eb' = c\eb$ and $\mu' = \mu/c$, and that the value of $c$ cannot be assessed directly. Therefore, if one is unaware of $c$, the analysis performed underestimates the transmissions by a factor of $c$ and overestimates the brightness by a factor of $1/c$. However, we can show by direct calculation that the predicted probability vector $\vector{P}$, and consequently the predicted $\gt$ for a given heralding rate, are independent of $c$. Therefore, in our results obtained in section~\ref{section:results}, we may have slightly overestimated the values of the brightness but we still predicted the correct value for the measured $\gt$.

For QKD with a HSPS, one can assess the security of the protocol against PNS attacks by knowing the value of $\mu$. The analysis we propose gives an overestimated value for $\mu$ when there are reasons to expect that spectral correlations and unmatched filtering are present. This is not detrimental to the security of QKD, as a sender unaware of this will, in the worst case, only overestimate the information available to an eavesdropper and shorten the key more than necessary through privacy amplification~\cite{BBR88,BBCM95}.

\section{Conclusion}
We developed a model exactly describing the detection statistics of a probabilistic source of photon pairs. From this model, we outlined a method from which the transmission of each photon channel, as well as the source's brightness, can be determined by measuring single and two-fold coincidence detection probabilities stemming from photons belonging to one pair. Then, we experimentally confirmed the method by demonstrating that the measured $\gt$ of a HSPS can be correctly predicted for any heralding probability. This allows one to quickly tune the brightness on demand as required to optimize the performance of entangled QKD, to assess the security of HSPS-based QKD or to optimize quantum repeater error rates and distances, all in the context of fluctuating experimental conditions such as photon channel transmissions. Finally, we showed that our model correctly reproduces the detection statistics even if the photons are spectrally and/or spatially correlated, and that this only leads to an overestimation of the brightness of the source. The simplicity of the proposed method makes it very attractive for the field of quantum communication in general.

\section*{Acknowledgments}
This work is supported by NSERC, iCORE, GDC, CFI, AAET, QuantumWorks, NATEQ, AIF and CIPI.
\ \\ \ \\
$^{\dag}$These authors contributed equally to this work.

\begin{thebibliography}{99}
\bibitem{thermalg2} For a Thermal distribution the $\gt$ is higher by a factor of $2$: $\gt = 2\mu(2-\eh)$. In the case of spectral and/or spatial correlations with a Poissonian source, as discussed in section~\ref{section:correlations}, $\gt = \mu(2/c - \eh)$.
\end{thebibliography}
\end{document}